\documentclass[aps,pre,showpacs]{revtex4}
\usepackage{graphicx}
\usepackage{epsfig}
\usepackage{amsmath}
\usepackage{graphics}
\begin{document}

\title{Asymptotic non-equilibrium steady state operators}

\author{M. F. Gelin}

\affiliation{Department of Chemistry, Technical University of Munich, Lichtenbergstrasse 4, D-85747 Garching, Germany}

\author{D. S. Kosov$^{1,2}$ }
\affiliation{
$^1$Department of Physics and  Center for Nonlinear Phenomena and Complex Systems,
Universit\'e Libre de Bruxelles, 
Campus Plaine, CP 231, 
Blvd du Triomphe, 
B-1050 Brussels, 
Belgium \\
$^2$Department of Chemistry and Biochemistry,
 University of Maryland, College Park, 20742, USA  }


\begin{abstract}
We present a method for the calculation of asymptotic operators for non-equilibrium steady-state
quantum  systems. The asymptotic steady-state operator is obtained by  averaging 
the corresponding operator in Heisenberg representation over  
infinitely long time. Several examples  are considered 
to demonstrate the utility of our method. The results obtained within our approach are 
compared to those obtained within the  Schwinger-Keldysh non-equilibrium Green's functions.
\end{abstract}

\maketitle

\section{Introduction}

If we place a finite quantum or classical system into the 
contact with several different macroscopic  thermal baths or  particle reservoirs
 and wait for some time, which is much longer than  typical relaxation time of the system,
the system  will reach a  non-equilibrium time-independent steady state.
Like an equilibrium represents stationary state of a closed system, non-equilibrium steady state is the time-inviriant state of an open system.
 The steady state is established by the delicate balance between irreversible processes and the driving forces produced by the reservoirs.
 Non-equilibrium steady-state systems are ubiquitous and their
theoretical description  has been a challenging fundamental problem for many years \cite{gaspard06}.  They are also of significant practical interest for various 
nanotechnological  and biological applications. 
Examples of steady-state non-equilibrium systems include
quantum contacts \cite{Agrait200381},  molecular motors \cite{andrieux:011906,kolomeisky,Mikhailov07}, 
molecular junctions and nanowires \cite{Nitzan03}, low dimensional heat conducting quantum and classical systems  \cite{dhar01,dhar:134301,segal:105901}. 
Recent  experiments on these systems have revealed a wealth of interesting new non-equilibrium phenomena such as non-diffusive heat transfer, negative differential resistance, stochastic switching  and hysteresis of electric current, and direct harnessing of 
thermal  fluctuations \cite{cahill:793,venkataraman:458,xu:1221}.

The understanding of the fundamental mechanisms  as well as the interpretation of these experimental observations require the development of new  theoretical methods for the description of non-equilibrium quantum systems.  Many theoretical approaches have been developed to deal with non-equilibrium steady-state systems, such as 
Keldysh-Schwinger Green's
 functions \cite{keldysh65,Caroli71,Jauho94,Meir92,dahnovsky:014104,dahnovsky:234111,rammer-book}, scattering theory based approaches \cite{diventra:979,Lang95},  different variations of Zubarev method of non-equilibrium statistical operator \cite{zubarev-book,hershfield:2134,tasaki06,gelin:011116,anders:066804}, and constrained current methods \cite{jaynes,PhysRevB.51.14421,kosov:6368,kosov:7165,bokes:165425,kosov:1,kosov-greer, painelli:155305}. 
Most of the  methods are based on  calculations of  non-equilibrium averages of  the operators of interest. Although average values, such as correlation functions, densities, currents, are very important and directly related to the measurable quantities, it is very useful to have explicit expressions for  asymptotic operators. Here we propose a general theoretical approach which directly defines and computes  asymptotic non-equilibrium steady-state operators.

\section{Asymptotic  operators}

Let us consider a quantum mechanical observable which corresponds to a Hermitian operator $A$.
We assume the existence of a unique steady state for our system. It implies the existence of the limit $t \rightarrow + \infty$ for observable quantities. It means that the asymptotic value of the  operator $ {A}$ in Heisenberg representation  is uniquely defined and can be computed as
\begin{eqnarray}
\overline{A}=\lim_{t \to +\infty}e^{iHt}  {A} e^{-iHt},
\label{limit}
\end{eqnarray}
where the Hamiltonian  $H$ includes the system Hamiltonian $H_S$,  the bath Hamiltonian $H_B$, and the system-bath interaction $H_{SB}$:
\begin{equation}
H= H_S + H_B + H_{SB}.
\label{totalh}
\end{equation}
We denote all thermal baths and particle reservoirs  collectively by $B$. The Hamiltonian $H$ 
is assumed to be time-independent.
We begin calculation of the limit (\ref{limit}) by the use of the following self-evident identity, 
which is valid for any time-dependent  operator $O(t)$:
\begin{equation}
O(t)=O(0) +\int_0^t dt' \frac{dO(t')}{dt'}. \label{O}
\end{equation}
Applying this identity  to eq.(\ref{limit}), we get
\begin{equation}
\overline{A}= A +\int_0^{+\infty} dt \frac{d}{dt} e^{i {H}t}  {A} e^{-i {H}t} .  
\label{current-1}
\end{equation}
The existence of the unique steady state implies the convergence of the integral in eq.(\ref{current-1}), therefore we can introduce the infinitesimal factor $\eta$:
\begin{equation}
\overline{A}= A + \lim_{\eta\rightarrow +0} \int_0^{+\infty} dt  e^{-\eta t} \frac{d}{dt}  e^{i {H}t}  {A} e^{-i {H}t} .
\label{current-2}
\end{equation}
This regularization of the  integral  is equivalent to the assumption that there exist some dissipation mechanisms in our system which we do not specify explicitly but which are efficient enough to lead to the  unique steady state.
The integration by parts of eq.(\ref{current-2}) yields
 \begin{equation}
\overline{A}= \lim_{\eta\rightarrow +0} \eta \int_0^{+\infty} dt  e^{- \eta t} e^{i {H}t}  {A} e^{-i {H}t}  .
\label{current-3}
\end{equation}
This expression for the non-equilibrium asymptotic operator is interesting physically since the integral is multiplied by the infinitesimal number $\eta$. 
Therefore, only singular terms proportional to $1/\eta$ give non-vanishing contribution from the integral to $\overline{A}$. 
Using   Abel's theorem \cite{zubarev-book}
\begin{equation}
\lim_{T\rightarrow \infty}\frac{1}{T} \int_0^T dt f(t)  =\lim_{\eta \rightarrow +0} \eta \int_0^{+\infty} dt  e^{- \eta t} f(t),
\label{abel}
\end{equation}
we see that eq.(\ref{current-3}) yields the part of the operator which is conserved after averaging over the infinitely long time.
Eq.(\ref{current-3}) is the definition of the  asymptotic non-equilibrium steady-state operator. It  will be the  starting point for our calculations. The definition (\ref{current-3}) has  been initially
 introduced by Kubo in his work on irreversible thermodynamics (so-called time invariant part of the operator)\cite{kubo59}  and Zubarev during his development of 
 non-equilibrium statistical operator method \cite{zubarev-book}. It  was used Grandy \cite{Grandy2} and Hershfield \cite{hershfield:2134} in their formulations of steady state non-equlibrium statistical mechanics.  Recently, Tasaki and Takahashi  also employed asymptotic operators (\ref{current-3}) to continue the advances of Zubarev method in application to transport in quantum junctions \cite{tasaki06}. The time invariant current defined via (\ref{abel}) was used by
 Bokes, Mera, and Godby as a term constrained by  the Lagrange multiplier in their development  of variational transport  theory\cite{bokes:165425}.
 In this paper, we develop a general practical method for computing asymptotic non-equilibrium operators.

To carry out our program it is convenient  to represent the Hamiltonian and operator  $A$ in the second quantization form.  We assume that the Hamiltonian and the operator $A$ are quadratic in creation and annihilation operators:
\begin{equation}
H= \sum_{ij} H_{ij} a^{\dagger}_i a_j,
\label{hqua}
\end{equation}
\begin{equation}
A= \sum_{ij} A_{ij} a^{\dagger}_i a_j.
\label{aqua}
\end{equation}
The creation $a^{\dagger}$ and annihilation $a$ operators  obey the standard commutation (for bosons)  and anticommutation (for fermions) relations. The methods developed in this paper are  applicable to both, Fermi and Bose, systems.
With the use of the following operator identity 
\begin{equation}
e^{i H t} A e^{-iHt}  =
\sum_{ij} \left( e^{i \mathbf{H} t} \mathbf{A} e^{-i \mathbf{H} t} \right)_{ij} a^{\dagger}_i a_j ,
\label{operator-relation}
\end{equation}
where  $\mathbf{H}$ and $\mathbf{A}$  are  matrices with matrix elements  $H_{ij}$  (\ref{hqua}) and   $A_{ij}$  (\ref{aqua}), 
the expression for the asymptotic operator (\ref{current-3}) becomes 
\begin{equation}
\overline{A}=
\lim_{\eta\rightarrow +0} 2 \eta \int_0^{+\infty} dt 
 \sum_{ij} \left[
 e^{-i(\omega-\mathbf{H} -i \eta)t} \mathbf{A} 
  e^{i(\omega-\mathbf{H} +i \eta)t}  \right]_{ij} a^{\dagger}_i a_j.
\label{current-4}
\end{equation}
This integral can further be transformed by inserting the delta function
 $\delta(t-t') =\frac{1}{2\pi} \int_{-\infty}^{+\infty} d\omega e^{i\omega (t-t')}$ 
 and integrating over $t$ and $t'$:
\begin{equation}
\overline{A}=
\lim_{\eta\rightarrow +0} \frac{\eta}{\pi} \int_{-\infty}^{+\infty} d\omega  
 \sum_{ij} \left[ \mathbf{G}^{*}(\omega)
\mathbf{A} 
 \mathbf{G}(\omega)  \right]_{ij} a^{\dagger}_i a_j,
\label{current-5}
\end{equation}
where $\mathbf{G}(\omega)= [\omega -\mathbf{H} + i \eta]^{-1}$ is  the total Green's function.  This is the general formula for the asymptotic steady-state non-equilibrium operator, which is applicable to any systems with quadratic Hamiltonians.

The indexes $i$ and $j$ run over the system $s$ and bath $b$ single particle states in eqs.(\ref{hqua}) and (\ref{aqua}).
Therefore the total Hamiltonian  matrix can be written in the following form
\begin{equation}
\mathbf{H}=\left[\begin{array}{cc}
\mathbf{H}_B & \mathbf{H}_{SB}  \\
\mathbf{H}_{BS} & \mathbf{H}_S  \\
 \end{array} \right] ,
\label{hsb}
\end{equation}
where  $\mathbf{H}_S$ is the matrix of the  system Hamiltonian, $\mathbf{H}_B$ is the matrix of the bath Hamiltonians, and $\mathbf{H}_{SB}$ is the matrix of the  system-bath interaction. 
Likewise, we can partition the matrix $\mathbf{A}$
\begin{equation}
\mathbf{A}=\left[\begin{array}{cc}
\mathbf{A}_B & \mathbf{A}_{SB}  \\
\mathbf{A}_{BS} & \mathbf{A}_S  \\
 \end{array} \right] ,
\label{asbe}
\end{equation}
and the Green's function
\begin{equation}
\mathbf{G}(\omega)= [\omega -\mathbf{H} + i \eta]^{-1}= \left[\begin{array}{cc}
\mathbf{G}_B(\omega) & \mathbf{G}_{SB}(\omega)  \\
\mathbf{G}_{BS}(\omega) & \mathbf{G}_S(\omega)  \\
 \end{array} \right].
\label{asb}
\end{equation}
The block matrix elements of the Green's function, which can be computed by the Frobenius formula,\cite{gantmacher} have the following form
\begin{equation}
\mathbf{G}_{S}({\omega}) =\left[ (\omega +i \eta) \mathbf{I}_S -\mathbf{H}_S - {\Sigma}_B(\omega)   \right]^{-1},
 \label{gmm}
\end{equation}
\begin{equation}
\mathbf{G}_{SB}({\omega}) = - \mathbf{G}_{S} (\omega) \mathbf{H}_{SB} \mathbf{g}_B (\omega),
\label{gbs}
\end{equation}
\begin{equation}
\mathbf{G}_{B}({\omega}) = \mathbf{g}_B (\omega) + 
\mathbf{g}_B (\omega) \mathbf{H}_{BS} \mathbf{G}_{S} (\omega) \mathbf{H}_{SB} \mathbf{g}_B (\omega).
\end{equation}
Here  we  introduced the  Green's functions for the baths
\begin{equation}
\mathbf{g}_{B}({\omega}) =\left[ (\omega +i \eta) \mathbf{I}_{B}-\mathbf{H}_{B}   \right]^{-1},
\end{equation}
 and the corresponding self-energies
 \begin{equation}
{\Sigma}_{B}({\omega}) = \mathbf{H}_{SB} \mathbf{g}_{B} (\omega) \mathbf{H}_{BS}.
\end{equation}
Note that the baths are assumed to be macroscopically large, so they have continuum energy levels. Therefore, when $\eta$ tends to zero, 
the Green's function of the baths $\mathbf{g}_{B}({\omega})  $ has a pole on the real energy axis. The Green's 
function $\mathbf{G}_{S}({\omega})$ is not singular in general, although special care should be taken to deal with the bound states \cite{dhar:085119}.

We consider two practically important choices of the operator $A$ at time $t=0$. First, $A$ is defined in the system  space only 
\begin{eqnarray}
\overline{A}_S &=& \lim_{\eta\rightarrow +0}
\frac{\eta}{\pi} \int_{-\infty}^{+\infty} d\omega  
 \sum_{ij} \left[ \mathbf{G}^{*}(\omega)
\mathbf{A}_S 
 \mathbf{G}(\omega)  \right]_{ij} a^{\dagger}_i a_j
 = \lim_{\eta\rightarrow +0} \frac{\eta}{\pi} \int_{-\infty}^{+\infty} d\omega  
 \sum_{bb'} \left[ \mathbf{G}^{*}_{BS}(\omega)
\mathbf{A}_S 
 \mathbf{G}_{SB} (\omega)  \right]_{bb'} a^{\dagger}_b a_{b'}.
\label{as}
\end{eqnarray}
Second, $A$ mixes the Fock spaces of the system and baths:
\begin{eqnarray}
\overline{A}_{SB}&=&
\lim_{\eta\rightarrow +0} \frac{\eta}{\pi} \int_{-\infty}^{+\infty} d\omega  
 \sum_{ij} \left[ \mathbf{G}^{*}(\omega)
\mathbf{A}_{SB} 
 \mathbf{G}(\omega)  \right]_{ij} a^{\dagger}_i a_j =  \lim_{\eta\rightarrow +0}\frac{\eta}{\pi} \int_{-\infty}^{+\infty} d\omega  
 \sum_{bs'} \left[ \mathbf{G}^{*}_{BS}(\omega)
\mathbf{A}_{SB} 
 \mathbf{G}_{BS} (\omega)  \right]_{bs'} a^{\dagger}_b a_{s'}
 \nonumber
 \\
 && +
 \lim_{\eta\rightarrow +0} \frac{\eta}{\pi} \int_{-\infty}^{+\infty} d\omega  
 \sum_{bb'} \left[ \mathbf{G}^{*}_{S}(\omega)
\mathbf{A}_{SB} 
 \mathbf{G}_{B} (\omega)  \right]_{bb'} a^{\dagger}_s a_{b'} 
 +
 \lim_{\eta\rightarrow +0} \frac{\eta}{\pi} \int_{-\infty}^{+\infty} d\omega  
 \sum_{bb'} \left[ \mathbf{G}^{*}_{BS}(\omega)
\mathbf{A}_{SB} 
 \mathbf{G}_{B} (\omega)  \right]_{bb'} a^{\dagger}_b a_{b'}. 
\label{asbb}
\end{eqnarray}
Here the indexes $b$ and $s$  refer to the single-particle states of the baths and the system, respectively. 
We have retained only those terms under the integrals which are
$\sim 1/((\omega -\varepsilon_{b})^2 +\eta^2) $.  Less singular terms give zero contributions upon the 
integration and multiplication by $\eta$. With an eye on the averaging of asymptotic operators over the initial equilibrium density matrix, which is the product  of equilibrium system and bath density matrices, the expression for $\overline{A}_{SB}$ can further be simplified:
\begin{equation}
\overline{A}_{SB} \approx    \lim_{\eta\rightarrow +0}  \frac{\eta}{\pi} \int_{-\infty}^{+\infty} d\omega  
 \sum_{bb'} \left[ \mathbf{G}^{*}_{BS}(\omega)
\mathbf{A}_{SB} 
 \mathbf{G}_{B} (\omega)  \right]_{bb'} a^{\dagger}_b a_{b'}. 
 \label{asb2}
\end{equation}

\section{Comparisons with non-equilibrium Green's function method}

In order to compare our approach with existing theoretical methods, we consider a finite quantum system with discrete energy levels and non-interacting fermions (quantum dot, atom, atomic cluster, molecule, or  atomic Fermi-gas in a trap) which is attached to two macroscopic reservoirs. Such kind of systems can be described by the tunneling Hamiltonian \cite{cohen62},
which has the form
\begin{equation}
 {H} =  {H}_0 +  {T}.
\end{equation}
Here $H_0$ is the Hamiltonian of the isolated reservoirs and the system,
\begin{equation}
 {H_0}= \sum_{b=l,r} \varepsilon_{b} a^{\dagger}_b a_b   +
 \sum_s \varepsilon_s a^{\dagger}_s a_s.
 \end{equation}
The  indexes $l$  and $r$ refer to the continuum single-particle states in  the left  and right reservoirs, respectively, and $s$ refers to the discrete single-particle states of the system. 
The tunneling interaction couples the reservoirs to the system,
\begin{equation}
 {T}=\sum_{b=l,r}  \sum_s t_{bs} (a^{\dagger}_b a_s + a^{\dagger}_s a_b) .
 \label{tunneling}
\end{equation}
We deal with fermions in this example, therefore the creation $a^{\dagger}$ and annihilation $a$ operators obey the standard anticommutation relations. We assume that  the system is isolated from the reservoirs  at $t<0$ .
Therefore the system is in equilibrium  at $t<0$ and it is described by the density matrix
\begin{equation}
 {\rho}_0 =\frac{1}{\mbox{Tr}[e^{-\beta( {H}_0 -\mu_L  {N}_L -\mu_R  {N}_R)}]} e^{-\beta( {H}_0 -\mu_L  {N}_L -\mu_R  {N}_R)}.\label{rho0}
\end{equation}
Here $ {N}_L$ ($ {N}_R$) are the number of particles operators for the left(right) reservoirs
and $ {\mu}_L$ ($ {\mu}_R$) are the corresponding chemical potentials.
Then at $t=0$ we turn on the reservoir-system coupling $T$.
 
First, we compute  the asymptotic single-particle density matrix for the system
\begin{equation}
\overline{n}_{ss'}= \lim_{t\rightarrow +\infty}
 e^{iHt} a^{\dagger}_{s'} a_s e^{-iHt}.
\label{nss-1}
\end{equation}
The use of formula (\ref{as}) gives
\begin{equation}
\overline{n}_{ss'} = \lim_{\eta\rightarrow +0}
\frac{\eta}{\pi} \int_{-\infty}^{+\infty} d\omega  
 \sum_{bb'  } \left[ \mathbf{G}_{BS}^{*}(\omega)\right]_{bs'}
\left[
 \mathbf{G}_{SB}(\omega)  \right]_{sb'} a^{\dagger}_b a_{b'},
\label{nss-2}
\end{equation}
where $b$ and $b'$ run over all single-particle states of the left and right reservoirs.
Substituting the explicit expressions for the Green's functions (\ref{gbs}), we get
\begin{equation}
\overline{n}_{ss'} = \lim_{\eta\rightarrow +0}
\frac{\eta}{\pi} \int_{-\infty}^{+\infty} d\omega  
 \sum_{bb'  } \left[
  \mathbf{g}^*_B (\omega)
  \mathbf{H}_{BS}
    \mathbf{G}^*_{S} (\omega)
 \right]_{bs'}
\left[  \mathbf{G}_{S} (\omega) \mathbf{H}_{SB} \mathbf{g}_B (\omega)
 \right]_{sb'} a^{\dagger}_b a_{b'}.
\label{nss-3}
\end{equation}
Averaging the asymptotic operator (\ref{nss-3}) over the equilibrium initial density matrix, we obtain the following expression for the steady-state single-particle density matrix:
\begin{eqnarray}
\mbox{Tr} \left\{ \rho_0 \overline{n}_{ss'}\right\}=
\frac{1}{2 \pi i} \int_{-\infty}^{+\infty}  d\omega  
\sum_{B=L,R}
 f(\omega - \mu_B)
 \left[  \mathbf{G}_{S}({\omega}) ({\Sigma}^*_B(\omega) - 
{\Sigma}_B(\omega) )
  \mathbf{G}_{S}^*({\omega})
 \right]_{ss'} ,
\label{integral-8}
\end{eqnarray}
$f(\omega)$ being the Fermi-Dirac occupation numbers.
The same equation is obtained by the method of non-equilibrium Green's functions
through the lesser Green's function $G^{<}$ \cite{arnold:174101}.

Now we calculate the  asymptotic current operator.
The current operator is defined by the continuity equation:
\begin{equation}
 {J} =\frac{d}{dt}  {N}_L=i[ {H}, {N}_L] 
=i\sum_{ls} T_{ls} (a^{\dagger}_l a_s - a^{\dagger}_s a_l),
\end{equation}
$N_L$ being the particle number operator for the left reservoir
\footnote{A heat flow through the system can be studied in exactly the same way as a particle current. We simply need to 
define the energy current operator as ${J_E} =\frac{d}{dt}  \sum_l \varepsilon_l a^{\dagger}_l a_l 
=i\sum_{ls}  \varepsilon_lT_{ls} (a^{\dagger}_l a_s - a^{\dagger}_s a_l)$. }.

We define the asymptotic value of the current operator as 
\begin{equation}
\overline{J}= \lim_{t\rightarrow + \infty} 
 e^{iHt}{J} e^{-iHt} .
\label{current-int}
\end{equation}
The use of eq.(\ref{asb2}) leads to
\begin{equation}
\overline{J} =     \lim_{\eta\rightarrow +0} \frac{\eta}{\pi} \int_{-\infty}^{+\infty} d\omega  
 \sum_{bb'} \left[ \mathbf{G}^{*}_{BS}(\omega)
\mathbf{J} 
 \mathbf{G}_{B} (\omega)  \right]_{bb'} a^{\dagger}_b a_{b'},
 \label{jbs}
\end{equation}
where the matrix, which represents the current, is 
\begin{equation}
\mathbf{J}=\left[\begin{array}{ccc}
0  & \mathbf{T}_{LN} & 0 \\
-\mathbf{T}_{NL} & 0 & 0 \\
0 & 0 & 0  \end{array} \right] \;.
\label{current-mat}
\end{equation}
Substituting the expressions for $G_B(\omega)$ and $G_{SB}(\omega)$ into eq. (\ref{jbs}), and averaging over the equilibrium initial density matrix (\ref{rho0}) results into the following expression  for the steady-state current:
\begin{equation}
\overline{J}= 
\frac{1}{2 \pi} \int_{-\infty}^{+\infty}  d\omega  (n(\omega-\mu_L) -n(\omega-\mu_R))
\mbox{Tr} \left[ ({\Sigma}_L(\omega) - {\Sigma}^*_L(\omega)) 
  \mathbf{G}_{NN}^*({\omega})( {\Sigma}_{R}({\omega}) 
-{\Sigma}^*_{R}({\omega}) )
 \mathbf{G}_{NN}({\omega})   \right] .
\label{landauer}
\end{equation} 
 This is the so-called Landauer formula, which is typically  obtained via 
the formalism of non-equilibrium Green's  functions \cite{Caroli71,Meir92}.

\section{Conclusions} 

We define a non-equilibrium steady state as an asymptotic state of a finite quantum system which is connected to several different macroscopic thermal baths or particle reservoirs. 
We present a  general method to compute asymptotic steady-state operators for such systems. 
The asymptotic operators can be computed exactly for the systems with  quadratic Hamiltonians. For the tunneling Hamiltonian we recover the standard results obtained by the non-equilibrium Green's function method, thereby demonstrating  equivalence of the two approaches. For example,  we derive Landauer formula by a direct calculation of the asymptotic steady-state non-equilibrium current operator.

\begin{acknowledgments}
This work has been supported by the Francqui Foundation,  American Chemical Society Petroleum Research Fund (44481-G6),  Deutsche Forschungsgemeinschaft (DFG)  through the DFG Cluster of Excellence ``Munich Centre of Advanced Photonics'' 
(www.munich-photonics.de).

\end{acknowledgments}

%\begin{thebibliography}{10}
\bibliographystyle{apsrev}
\bibliography{/Users/dkosov/Documents/BIBLIOGRAPHY/tddft-moletronics}

\begin{thebibliography}{40}
\expandafter\ifx\csname natexlab\endcsname\relax\def\natexlab#1{#1}\fi
\expandafter\ifx\csname bibnamefont\endcsname\relax
  \def\bibnamefont#1{#1}\fi
\expandafter\ifx\csname bibfnamefont\endcsname\relax
  \def\bibfnamefont#1{#1}\fi
\expandafter\ifx\csname citenamefont\endcsname\relax
  \def\citenamefont#1{#1}\fi
\expandafter\ifx\csname url\endcsname\relax
  \def\url#1{\texttt{#1}}\fi
\expandafter\ifx\csname urlprefix\endcsname\relax\def\urlprefix{URL }\fi
\providecommand{\bibinfo}[2]{#2}
\providecommand{\eprint}[2][]{\url{#2}}

\bibitem[{\citenamefont{Gaspard}(2006)}]{gaspard06}
\bibinfo{author}{\bibfnamefont{P.}~\bibnamefont{Gaspard}},
  \bibinfo{journal}{Progress of Theoretical Physics Supplement}
  \textbf{\bibinfo{volume}{165}} (\bibinfo{year}{2006}).

\bibitem[{\citenamefont{Agrait et~al.}(2003)\citenamefont{Agrait, Yeyati, and
  van Ruitenbeek}}]{Agrait200381}
\bibinfo{author}{\bibfnamefont{N.}~\bibnamefont{Agrait}},
  \bibinfo{author}{\bibfnamefont{A.~L.} \bibnamefont{Yeyati}},
  \bibnamefont{and} \bibinfo{author}{\bibfnamefont{J.~M.} \bibnamefont{van
  Ruitenbeek}}, \bibinfo{journal}{Physics Reports}
  \textbf{\bibinfo{volume}{377}}, \bibinfo{pages}{81 } (\bibinfo{year}{2003}),
  ISSN \bibinfo{issn}{0370-1573}.

\bibitem[{\citenamefont{Andrieux and Gaspard}(2006)}]{andrieux:011906}
\bibinfo{author}{\bibfnamefont{D.}~\bibnamefont{Andrieux}} \bibnamefont{and}
  \bibinfo{author}{\bibfnamefont{P.}~\bibnamefont{Gaspard}},
  \bibinfo{journal}{Phys. Rev. E} \textbf{\bibinfo{volume}{74}},
  \bibinfo{eid}{011906} (\bibinfo{year}{2006}).

\bibitem[{\citenamefont{Kolomeisky and Fisher}(2007)}]{kolomeisky}
\bibinfo{author}{\bibfnamefont{A.~B.} \bibnamefont{Kolomeisky}}
  \bibnamefont{and} \bibinfo{author}{\bibfnamefont{M.~E.}
  \bibnamefont{Fisher}}, \bibinfo{journal}{Annual Review of Physical Chemistry}
  \textbf{\bibinfo{volume}{58}}, \bibinfo{pages}{675} (\bibinfo{year}{2007}).

\bibitem[{\citenamefont{Togashi and Mikhailov}(2007)}]{Mikhailov07}
\bibinfo{author}{\bibfnamefont{Y.}~\bibnamefont{Togashi}} \bibnamefont{and}
  \bibinfo{author}{\bibfnamefont{A.~S.} \bibnamefont{Mikhailov}},
  \bibinfo{journal}{Proc. Natl. Acad. Sci. U.S.A.}
  \textbf{\bibinfo{volume}{104}} (\bibinfo{year}{2007}).

\bibitem[{\citenamefont{Nitzan and Ratner}(2003)}]{Nitzan03}
\bibinfo{author}{\bibfnamefont{A.}~\bibnamefont{Nitzan}} \bibnamefont{and}
  \bibinfo{author}{\bibfnamefont{M.~A.} \bibnamefont{Ratner}},
  \bibinfo{journal}{Science} \textbf{\bibinfo{volume}{300}},
  \bibinfo{pages}{1384} (\bibinfo{year}{2003}).

\bibitem[{\citenamefont{Dhar}(2001)}]{dhar01}
\bibinfo{author}{\bibfnamefont{A.}~\bibnamefont{Dhar}}, \bibinfo{journal}{Phys.
  Rev. Lett.} \textbf{\bibinfo{volume}{86}}, \bibinfo{pages}{5882}
  (\bibinfo{year}{2001}).

\bibitem[{\citenamefont{Dhar and Lebowitz}(2008)}]{dhar:134301}
\bibinfo{author}{\bibfnamefont{A.}~\bibnamefont{Dhar}} \bibnamefont{and}
  \bibinfo{author}{\bibfnamefont{J.~L.} \bibnamefont{Lebowitz}},
  \bibinfo{journal}{Phys. Rev. Lett.} \textbf{\bibinfo{volume}{100}},
  \bibinfo{eid}{134301} (\bibinfo{year}{2008}).

\bibitem[{\citenamefont{Segal}(2008)}]{segal:105901}
\bibinfo{author}{\bibfnamefont{D.}~\bibnamefont{Segal}},
  \bibinfo{journal}{Phys. Rev. Lett.} \textbf{\bibinfo{volume}{100}},
  \bibinfo{eid}{105901} (\bibinfo{year}{2008}).

\bibitem[{\citenamefont{Cahill et~al.}(2003)\citenamefont{Cahill, Ford,
  Goodson, Mahan, Majumdar, Maris, Merlin, and Phillpot}}]{cahill:793}
\bibinfo{author}{\bibfnamefont{D.~G.} \bibnamefont{Cahill}},
  \bibinfo{author}{\bibfnamefont{W.~K.} \bibnamefont{Ford}},
  \bibinfo{author}{\bibfnamefont{K.~E.} \bibnamefont{Goodson}},
  \bibinfo{author}{\bibfnamefont{G.~D.} \bibnamefont{Mahan}},
  \bibinfo{author}{\bibfnamefont{A.}~\bibnamefont{Majumdar}},
  \bibinfo{author}{\bibfnamefont{H.~J.} \bibnamefont{Maris}},
  \bibinfo{author}{\bibfnamefont{R.}~\bibnamefont{Merlin}}, \bibnamefont{and}
  \bibinfo{author}{\bibfnamefont{S.~R.} \bibnamefont{Phillpot}},
  \bibinfo{journal}{J. App. Phys.} \textbf{\bibinfo{volume}{93}},
  \bibinfo{pages}{793} (\bibinfo{year}{2003}).

\bibitem[{\citenamefont{Venkataraman et~al.}(2006)\citenamefont{Venkataraman,
  Klare, Tam, Nuckolls, Hybertsen, and Steigerwald}}]{venkataraman:458}
\bibinfo{author}{\bibfnamefont{L.}~\bibnamefont{Venkataraman}},
  \bibinfo{author}{\bibfnamefont{J.}~\bibnamefont{Klare}},
  \bibinfo{author}{\bibfnamefont{I.}~\bibnamefont{Tam}},
  \bibinfo{author}{\bibfnamefont{C.}~\bibnamefont{Nuckolls}},
  \bibinfo{author}{\bibfnamefont{M.}~\bibnamefont{Hybertsen}},
  \bibnamefont{and}
  \bibinfo{author}{\bibfnamefont{M.}~\bibnamefont{Steigerwald}},
  \bibinfo{journal}{Nano Letters} \textbf{\bibinfo{volume}{6}},
  \bibinfo{pages}{458} (\bibinfo{year}{2006}).

\bibitem[{\citenamefont{Xu and Tao}(2003)}]{xu:1221}
\bibinfo{author}{\bibfnamefont{B.}~\bibnamefont{Xu}} \bibnamefont{and}
  \bibinfo{author}{\bibfnamefont{N.~J.} \bibnamefont{Tao}},
  \bibinfo{journal}{Science} \textbf{\bibinfo{volume}{301}},
  \bibinfo{pages}{1221} (\bibinfo{year}{2003}).

\bibitem[{\citenamefont{Keldysh}(1965)}]{keldysh65}
\bibinfo{author}{\bibfnamefont{L.~V.} \bibnamefont{Keldysh}},
  \bibinfo{journal}{Sov. Phys. JETP} \textbf{\bibinfo{volume}{20}},
  \bibinfo{pages}{1018} (\bibinfo{year}{1965}).

\bibitem[{\citenamefont{Caroli et~al.}(1971)\citenamefont{Caroli, Combesco,
  Nozieres, and Saintjam}}]{Caroli71}
\bibinfo{author}{\bibfnamefont{C.}~\bibnamefont{Caroli}},
  \bibinfo{author}{\bibfnamefont{R.}~\bibnamefont{Combesco}},
  \bibinfo{author}{\bibfnamefont{P.}~\bibnamefont{Nozieres}}, \bibnamefont{and}
  \bibinfo{author}{\bibfnamefont{D.}~\bibnamefont{Saintjam}},
  \bibinfo{journal}{J. Phys. C} \textbf{\bibinfo{volume}{4}},
  \bibinfo{pages}{916} (\bibinfo{year}{1971}).

\bibitem[{\citenamefont{Jauho et~al.}(1994)\citenamefont{Jauho, Wingreen, and
  Meir}}]{Jauho94}
\bibinfo{author}{\bibfnamefont{A.}~\bibnamefont{Jauho}},
  \bibinfo{author}{\bibfnamefont{N.}~\bibnamefont{Wingreen}}, \bibnamefont{and}
  \bibinfo{author}{\bibfnamefont{Y.}~\bibnamefont{Meir}},
  \bibinfo{journal}{Phys. Rev. B} \textbf{\bibinfo{volume}{50}},
  \bibinfo{pages}{5528} (\bibinfo{year}{1994}).

\bibitem[{\citenamefont{Meir and Wingreen}(1992)}]{Meir92}
\bibinfo{author}{\bibfnamefont{Y.}~\bibnamefont{Meir}} \bibnamefont{and}
  \bibinfo{author}{\bibfnamefont{N.}~\bibnamefont{Wingreen}},
  \bibinfo{journal}{Phys. Rev. Lett.} \textbf{\bibinfo{volume}{68}},
  \bibinfo{pages}{2512} (\bibinfo{year}{1992}).

\bibitem[{\citenamefont{Dahnovsky}(2007{\natexlab{a}})}]{dahnovsky:014104}
\bibinfo{author}{\bibfnamefont{Y.}~\bibnamefont{Dahnovsky}},
  \bibinfo{journal}{J. Chem. Phys.} \textbf{\bibinfo{volume}{127}},
  \bibinfo{eid}{014104} (\bibinfo{year}{2007}{\natexlab{a}}).

\bibitem[{\citenamefont{Dahnovsky}(2007{\natexlab{b}})}]{dahnovsky:234111}
\bibinfo{author}{\bibfnamefont{Y.}~\bibnamefont{Dahnovsky}},
  \bibinfo{journal}{J. Chem. Phys.} \textbf{\bibinfo{volume}{126}},
  \bibinfo{eid}{234111} (\bibinfo{year}{2007}{\natexlab{b}}).

\bibitem[{\citenamefont{Rammer}(2007)}]{rammer-book}
\bibinfo{author}{\bibfnamefont{J.}~\bibnamefont{Rammer}},
  \emph{\bibinfo{title}{Quantum Field Theory of Non-equilibrium States}}
  (\bibinfo{publisher}{Cambridge University Press}, \bibinfo{year}{2007}).

\bibitem[{\citenamefont{Di~Ventra et~al.}(2000)\citenamefont{Di~Ventra,
  Pantelides, and Lang}}]{diventra:979}
\bibinfo{author}{\bibfnamefont{M.}~\bibnamefont{Di~Ventra}},
  \bibinfo{author}{\bibfnamefont{S.~T.} \bibnamefont{Pantelides}},
  \bibnamefont{and} \bibinfo{author}{\bibfnamefont{N.~D.} \bibnamefont{Lang}},
  \bibinfo{journal}{Phys. Rev. Lett.} \textbf{\bibinfo{volume}{84}},
  \bibinfo{pages}{979} (\bibinfo{year}{2000}).

\bibitem[{\citenamefont{Lang}(1995)}]{Lang95}
\bibinfo{author}{\bibfnamefont{N.~D.} \bibnamefont{Lang}},
  \bibinfo{journal}{Phys. Rev. B} \textbf{\bibinfo{volume}{52}},
  \bibinfo{pages}{5335} (\bibinfo{year}{1995}).

\bibitem[{\citenamefont{Zubarev}(1974)}]{zubarev-book}
\bibinfo{author}{\bibfnamefont{D.~N.} \bibnamefont{Zubarev}},
  \emph{\bibinfo{title}{Nonequilibrium Statistical Thermodynamics}}
  (\bibinfo{publisher}{Consultants Bureau}, \bibinfo{year}{1974}).

\bibitem[{\citenamefont{Hershfield}(1993)}]{hershfield:2134}
\bibinfo{author}{\bibfnamefont{S.}~\bibnamefont{Hershfield}},
  \bibinfo{journal}{Phys. Rev. Lett.} \textbf{\bibinfo{volume}{70}},
  \bibinfo{pages}{2134} (\bibinfo{year}{1993}).

\bibitem[{\citenamefont{Tasaki and Takahashi}(2006)}]{tasaki06}
\bibinfo{author}{\bibfnamefont{S.}~\bibnamefont{Tasaki}} \bibnamefont{and}
  \bibinfo{author}{\bibfnamefont{J.}~\bibnamefont{Takahashi}},
  \bibinfo{journal}{Progress of Theoretical Physics Supplement}
  \textbf{\bibinfo{volume}{165}}, \bibinfo{pages}{57} (\bibinfo{year}{2006}).

\bibitem[{\citenamefont{Gelin and Kosov}(2008)}]{gelin:011116}
\bibinfo{author}{\bibfnamefont{M.~F.} \bibnamefont{Gelin}} \bibnamefont{and}
  \bibinfo{author}{\bibfnamefont{D.~S.} \bibnamefont{Kosov}},
  \bibinfo{journal}{Phys. Rev. E} \textbf{\bibinfo{volume}{78}},
  \bibinfo{eid}{011116} (\bibinfo{year}{2008}).

\bibitem[{\citenamefont{Anders}(2008)}]{anders:066804}
\bibinfo{author}{\bibfnamefont{F.~B.} \bibnamefont{Anders}},
  \bibinfo{journal}{Phys. Rev. Lett.} \textbf{\bibinfo{volume}{101}},
  \bibinfo{eid}{066804} (\bibinfo{year}{2008}).

\bibitem[{\citenamefont{Jaynes}(1957)}]{jaynes}
\bibinfo{author}{\bibfnamefont{E.~T.} \bibnamefont{Jaynes}},
  \bibinfo{journal}{Phys. Rev.} \textbf{\bibinfo{volume}{106}},
  \bibinfo{pages}{620} (\bibinfo{year}{1957}).

\bibitem[{\citenamefont{Johnson and Heinonen}(1995)}]{PhysRevB.51.14421}
\bibinfo{author}{\bibfnamefont{M.~D.} \bibnamefont{Johnson}} \bibnamefont{and}
  \bibinfo{author}{\bibfnamefont{O.}~\bibnamefont{Heinonen}},
  \bibinfo{journal}{Phys. Rev. B} \textbf{\bibinfo{volume}{51}},
  \bibinfo{pages}{14421} (\bibinfo{year}{1995}).

\bibitem[{\citenamefont{Kosov}(2002)}]{kosov:6368}
\bibinfo{author}{\bibfnamefont{D.~S.} \bibnamefont{Kosov}},
  \bibinfo{journal}{J. Chem. Phys.} \textbf{\bibinfo{volume}{116}},
  \bibinfo{pages}{6368} (\bibinfo{year}{2002}).

\bibitem[{\citenamefont{Kosov}(2004)}]{kosov:7165}
\bibinfo{author}{\bibfnamefont{D.~S.} \bibnamefont{Kosov}},
  \bibinfo{journal}{J. Chem. Phys.} \textbf{\bibinfo{volume}{120}},
  \bibinfo{pages}{7165} (\bibinfo{year}{2004}).

\bibitem[{\citenamefont{Bokes et~al.}(2005)\citenamefont{Bokes, Mera, and
  Godby}}]{bokes:165425}
\bibinfo{author}{\bibfnamefont{P.}~\bibnamefont{Bokes}},
  \bibinfo{author}{\bibfnamefont{H.}~\bibnamefont{Mera}}, \bibnamefont{and}
  \bibinfo{author}{\bibfnamefont{R.~W.} \bibnamefont{Godby}},
  \bibinfo{journal}{Phys. Rev. B} \textbf{\bibinfo{volume}{72}},
  \bibinfo{eid}{165425} (\bibinfo{year}{2005}).

\bibitem[{\citenamefont{Kosov}(2003)}]{kosov:1}
\bibinfo{author}{\bibfnamefont{D.~S.} \bibnamefont{Kosov}},
  \bibinfo{journal}{J. Chem. Phys.} \textbf{\bibinfo{volume}{119}},
  \bibinfo{pages}{1} (\bibinfo{year}{2003}).

\bibitem[{\citenamefont{Kosov and Greer}(2001)}]{kosov-greer}
\bibinfo{author}{\bibfnamefont{D.~S.} \bibnamefont{Kosov}} \bibnamefont{and}
  \bibinfo{author}{\bibfnamefont{J.~C.} \bibnamefont{Greer}},
  \bibinfo{journal}{Phys. Lett. A} \textbf{\bibinfo{volume}{291}},
  \bibinfo{pages}{45} (\bibinfo{year}{2001}).

\bibitem[{\citenamefont{Painelli}(2006)}]{painelli:155305}
\bibinfo{author}{\bibfnamefont{A.}~\bibnamefont{Painelli}},
  \bibinfo{journal}{Phys. Rev. B} \textbf{\bibinfo{volume}{74}},
  \bibinfo{eid}{155305} (\bibinfo{year}{2006}).

\bibitem[{\citenamefont{Kubo}(1959)}]{kubo59}
\bibinfo{author}{\bibfnamefont{R.}~\bibnamefont{Kubo}},
  \emph{\bibinfo{title}{Lectures in Theoretical Physics, ed. W. Britten}}
  (\bibinfo{publisher}{New York, Interscience}, \bibinfo{year}{1959}),
  vol.~\bibinfo{volume}{1}, p. \bibinfo{pages}{120}.

\bibitem[{\citenamefont{Grandy}(1988)}]{Grandy2}
\bibinfo{author}{\bibfnamefont{W.}~\bibnamefont{Grandy}},
  \emph{\bibinfo{title}{Foundations of Statistical Mechanics : Volume II:
  Nonequilibrium Phenomena (Fundamental Theories of Physics)}}
  (\bibinfo{publisher}{Springer}, \bibinfo{year}{1988}).

\bibitem[{\citenamefont{Gantmacher}(2005)}]{gantmacher}
\bibinfo{author}{\bibfnamefont{F.~R.} \bibnamefont{Gantmacher}},
  \emph{\bibinfo{title}{Applications of the theory of matrices}}
  (\bibinfo{publisher}{Dover}, \bibinfo{year}{2005}).

\bibitem[{\citenamefont{Dhar and Sen}(2006)}]{dhar:085119}
\bibinfo{author}{\bibfnamefont{A.}~\bibnamefont{Dhar}} \bibnamefont{and}
  \bibinfo{author}{\bibfnamefont{D.}~\bibnamefont{Sen}},
  \bibinfo{journal}{Phys. Rev. B} \textbf{\bibinfo{volume}{73}},
  \bibinfo{eid}{085119} (\bibinfo{year}{2006}).

\bibitem[{\citenamefont{Cohen et~al.}(1962)\citenamefont{Cohen, Falicov, and
  Phillips}}]{cohen62}
\bibinfo{author}{\bibfnamefont{M.~H.} \bibnamefont{Cohen}},
  \bibinfo{author}{\bibfnamefont{L.~M.} \bibnamefont{Falicov}},
  \bibnamefont{and} \bibinfo{author}{\bibfnamefont{J.~C.}
  \bibnamefont{Phillips}}, \bibinfo{journal}{Phys. Rev. Lett.}
  \textbf{\bibinfo{volume}{8}}, \bibinfo{pages}{316} (\bibinfo{year}{1962}).

\bibitem[{\citenamefont{Arnold et~al.}(2007)\citenamefont{Arnold, Weigend, and
  Evers}}]{arnold:174101}
\bibinfo{author}{\bibfnamefont{A.}~\bibnamefont{Arnold}},
  \bibinfo{author}{\bibfnamefont{F.}~\bibnamefont{Weigend}}, \bibnamefont{and}
  \bibinfo{author}{\bibfnamefont{F.}~\bibnamefont{Evers}}, \bibinfo{journal}{J.
  Chem. Phys.} \textbf{\bibinfo{volume}{126}}, \bibinfo{eid}{174101}
  (\bibinfo{year}{2007}).

\end{thebibliography}
%\bibliography{tddft-moletronics}
%\end{thebibliography}

\end{document}